
%
%
%
\magnification = \magstephalf
\font\smrm = cmr8
\font\bgbf = cmbx12
\font\ssbf = cmssbx10
\font\sstn = cmss10
\font\smit = cmti8
\def\sy{\scriptstyle}
\def\ssy{\scriptscriptstyle}
\def\veps{\varepsilon}
\def\ovwp{\overwithdelims ()}
\def\slsh#1{#1\!\!\! /}
\rightline{}
\bigskip
\bigskip
\bigskip
\centerline{\bgbf Ultramicro Black Holes and Finiteness of the Electromagnetic}
\centerline{\bgbf Contribution to the Electron Mass}
\medskip
\smallskip
\centerline{\sstn S. K. Kauffmann}
\centerline{\smit 750 Gonzalez Drive, Apt. 6D, San Francisco, CA 94132-2208}
\medskip
\smallskip
{\baselineskip = .85\baselineskip
\smrm It is argued that the nonintegrably singular energy density of the
electron's electromagnetic field (in both the classical point-charge model
and quantum electrodynamics) must entail very strong self-gravitational
effects, which, in turn, through black hole phenomena at finite radii, might
conceivably cut off the otherwise infinite electromagnetic contribution to
the electron's mass.  The general-relativistic equations for static,
spherically symmetric stellar structure are specialized to treat static,
spherically symmetric, nonnegative localized energy densities which may
exhibit nonintegrable singularities as ${\sy r \rightarrow 0}$.
The total mass (including the self-gravitational effects) of such a
system hinges on a constant of integration, which is normally put to zero,
the correct value for the weak-gravitational Newtonian limit, but which, when
gravitational effects become sufficiently strong, may need to be set to a
sufficiently large positive value to avoid singularities in physically inter%
pretable quantities derived from the metric tensor.  A consistent, continuous
procedure for choosing this constant, which puts it as close to the Newtonian
value of zero as is physically permissible, is adopted.  When it comes out
larger than zero, the system has a black hole, with positive Schwarzschild
radius ${\sy r_{\ssy s}}$.  The total mass of the system may be expressed as
that of the black hole, ${\sy r_{\ssy s}/(2G)}$, plus the integral over the
region outside the black hole of the original energy density.  In many situa%
tions, including the electromagnetic ones of interest here, ${\sy r_{\ssy s}}$
is just the radius where ${\sy 4 \pi r^{\ssy 2}}$ times the original
energy density---the energy per radial distance---first attains the limiting
value ${\sy (2G)^{\ssy -1}}$, at which value it then, in effect, ``freezes''
within the black hole.  Application of these results to the electron's
electromagnetic field energy density produces, for the classical point-charge
model, an electrostatic mass contribution which is many orders of magnitude
larger than the electron's measured mass.  For quantum electrodynamics, howev%
er, the result is an electromagnetic mass contribution which is approximately
equal to the electron's bare mass---thus about half of its measured mass.\par}
\normalbaselines
\centerline{\vbox{\hrule width 1.5in}}
\medskip
\centerline{{\bf I.}\enspace{\ssbf Introduction and Discussion of Results}}
\smallskip
Notwithstanding the great success of the renormalization program in quantum
electrodynamics, the electromagnetic contribution to the electron mass can%
not be related to the bare mass and charge parameters of the theory, as it
comes out infinite.  Such an infinity is already present in the simple clas%
sical point-charge model of the electron, making the problem a venerable one.
Indeed, the quantized theory improves the situation to a considerable extent,
as virtual pair creation in effect endows the electron with some spatial ex%
tension, such as a bit of smearing of its effective charge distribution at
distances smaller than its Compton wavelength [1].  This quantum structure
considerably ameliorates the degree of the divergence of the electromagnetic
mass contribution (from linear to logarithmic) but fails to eliminate it.

Of course, this particular difficulty with the electromagnetic contribution
to the electron mass is by no means the only example of apparently well-ground%
ed physical theory producing bewildering infinities.  The frequency-divergent
blackbody radiation spectrum obtained classically by Rayleigh and Jeans led
Planck to essay his quantum hypothesis.  Lorentz's classical demonstration
that the Rutherford atomic model must result in infinite radiation and instab%
ility as the elecron spirals inward toward the nucleus motivated Bohr to
enlarge Planck's quantum hypothesis to embrace atomic structure.  On a less
exalted level, the infinite total cross section for Coulomb scattering finds
its resolution in the observation that the real world always provides other
charged matter to shield the Coulomb potential at some finite distance.  All
of these successful resolutions of theoretical infinities were due to
recognition that the theoretical framework which produced them involved some
type of critical overidealization of the problem being treated---the neglect
of certain physical effects which lay {\it outside\/} the utilized theoretical
framework and happened to be crucial to the problem which generated the
infinity.  Thus in retrospect we can assert that the Rayleigh-Jeans and
Lorentz classical divergences are due to a crucial neglect of quantum effects,
while the infinite Coulomb cross section arises from neglecting the inevitable
eventual shielding of any Coulomb potential.  These ``infinity-killing''
precedents unequivocably suggest that the resolution of the infinite electro%
magnetic contribution to the electron mass should be sought {\it outside\/}
the context of the classical or quantum electromagnetic theory which produces
it.  In particular, the renormalization program, which operates entirely
within the context of quantum electrodynamics, cannot itself be expected to
fully resolve the infinity problems of that theory---indeed, although it
very usefully ``quarantines'' the infinities into just a few locations
such as the electromagnetic contribution to the electron mass, it does
{\it not\/} banish them.

We are thus now compelled to ask what nonelectromagnetic phenomena might
place a finite limit on the electromagnetic contribution to the electron
mass.  For the strong and weak nuclear forces, no obvious mechanism which
could accomplish this suggests itself.  Indeed, these forces are thought to
be described by renormalizable gauge theories with divergences similar to
those of quantum electrodynamics.  Thus, rather than curing the infinite
electromagnetic contribution to the electron mass, the theories of the strong
and weak forces would appear to contribute additional infinities of their
own.  The gravitational force, as it exists between two electrons, is so
utterly minute relative to the corresponding electrostatic force that con%
ventional thinking dismisses any significant role for it in quantum
electrodynamics without further consideration.  However, this ostensible
logic overlooks the very infinity in the electromagnetic contribution to the
electron mass with which we are wrestling here!  A {\it mass\/} contribution
which purely electromagnetic theory says is {\it infinite\/} could hardly fail
to overwhelmingly engage self-gravitational effects.  Moreover, bringing grav%
ity into the picture lends encouragement that the infinity may be resolv%
able because (1) gravity contributes negative potential energy which could
conceivably tend in the direction of cancelling the positive electromagnet%
ic field energy, and (2) very strong stimulation of the gravitational force
is known to produce finite Schwarzschild radius black-hole effects which
could very possibly serve to cut off the infinity in the electromagnetic
field contribution to the electron mass (the mass of a black hole isn't in%
finite; it is essentially determined by (proportional to) its Schwarzschild
radius).  The thinking which we essay here may strike the reader as unorth%
odox, but it is in fact profoundly conservative is spirit.  We take very
seriously the theoretical electromagnetic field contribution to the electron
mass (does the physicist exist who would make light of the result of a cal%
culation in electromagnetic theory?), and then ask whether that result
{\it itself\/} implies that known nonelectromagnetic phenomena must figure
prominently in the total physical picture.  For self-gravitational effects
the answer is overpoweringly in the affirmative.

At any ``ordinary'' distance from the electron, the energy density of its
electromagnetic field as calculated from purely electromagnetic theory
must be extremely accurate; it is only at extraordinarily short distances
that self-gravitational effects can significantly modify this electromag%
netic field energy density.  In this regard gravitational resolution of
the electron's infinite electromagnetic mass contribution would parallel
the resolution of the other infinities mentioned earlier.  The Rayleigh-%
Jeans blackbody spectrum is adequate at low frequencies; the Planck quantum
modification only makes a difference (albeit an ``infinity-killing'' one) at
sufficiently high frequencies.  The classical Lorentz model for the elec%
tron radiatively spiraling in toward the nucleus is accurate at large radii,
with the ``infinity-killing'' energy level quantization only entering promin%
ently at sufficiently small radii.  The differential cross section for Cou%
lomb scattering (which leads to a divergent total cross section) is accurate
for impact parameters up to where shielding of the Coulomb potential sets in.
An electromagnetic field energy density for the electron which is accurately
given by purely electromagnetic theory down to radii small enough to provoke
significant self-gravitational effects is thus deeply consonant in spirit
with the above ``infinity-killing'' precedents of the past.

In Section II we specialize the general-relativistic equations for static,
spherically symmetric stellar structure [2] to treat the self-gravitational
effects flowing from an arbitrary static, spherically symmetric, nonnegative
localized energy density which has been specified in a ``world'' where $G = 0$
(e.g., a static field energy density calculated in the quantum electrody%
namicist's ``world'' of flat-space electromagnetic theory).  We give particu%
lar attention to the case where this $G = 0$ static, spherically symmetric,
nonnegative localized enegy density tends toward a nonintegrable singularity
at the origin, as is the case for the electomagnetic field energy density of
the electron, and discuss locating (when $G$ is ``switched on'') the resulting
black-hole Schwarzschild radius and obtaining the now finite total energy.

For many situations, an adequate rendition of our results for the black-hole
Schwarzschild radius in such static, spherically symmetric cases is that it
is located where the $G = 0$ energy per radial distance (namely $4\pi r^2$
times the $G = 0$ energy density at r) attains the the limiting value $(2G)^%
{-1}$ (we use units where the speed of light $c = 1$).  An effective energy
(or mass) contribution from this black hole may be obtained directly from
this radius by using the familiar relation that a black hole's mass equals
its Schwarzschild radius divided by $(2G)$.  The remaining effective mass
contribution, which may be thought of as coming from that part of the energy
density which lies outside the Schwarzschild radius, can be calculated by
simply integrating the $G = 0$ energy density over this external region.
This simple, intuitively appealing procedure is fully adequate for obtaining
the electromagnetic contribution to the electron's mass.

In Section III we apply this procedure to the classical point-charge model
of the electron; the result for the electrostatic mass contribution is un%
acceptably large---many orders of magnitude larger than the measured elec%
tron mass.  We then extract an effective quantum electromagnetic field
energy density in ordinary space from the lowest order Feynman diagram
for the electromagnetic field contribution to the electron's mass [3].
This effective quantum electromagnetic field energy density is less singu%
lar as $r\rightarrow 0$ than is that of the classical point-charge model,
but is still not quite integrable there (the $G = 0$ effective total
electromagnetic energy comes out logarithmically divergent due to this
singularity).  In addition, it differs, by a constant multiplicative fac%
tor whose value is smaller than unity, from the classical point-charge
model in the limit $r\rightarrow \infty$ as well.  This large r deviation
represents a finite charge renormalization effect which arises from par%
tial shielding of the bare charge by the spontaneous (out of the vacuum)
production of a virtual pair and photon---an intrinsic aspect of this
Feynman diagram (this spontaneously produced virtual pair is responsible
as well for the above-described amelioration of the singularity as
$r\rightarrow 0$, relative to the classical point-charge case).  Use of
this effective quantum electromagnetic energy density yields a much smal%
ler Schwarzschild radius than does the classical point-charge model, and,
after taking account of the accompanying finite (but not negligible!)
charge renormalization, yields an electromagnetic mass contribution which
is approximately equal to the bare mass (i.e., approximately half of
the measured mass).  This quantum electrodynamics cum gravity result is
certainly not implausible, and is a pleasantly far cry from the unresolved
infinite electromagnetic mass contribution of the renormalization program.

One can question the appropriateness of combining quantum electrodynamics
with a completely classical approach to the consequent gravitational self-%
energy in the manner just described, but, practically speaking, there is no
viable alternative at this time.  One wouldn't know how to tackle quantized
gravitational theory other than by perturbation expansions in nonnegative
powers of $G$, which is hopelessly inappropriate to the ultrastrong gravi%
tational forces involved with black holes---the domain in which the above-%
described physics lies.  We can take some comfort from the thought that for
such very strong gravitational forces in a static situation, the classical
limit may very well be an adequate description.

In Section IV we entertain the speculation that gravitation can as well
resolve the remaining infinities of the quantum electrodynamics renormal%
ization program.  For the lowest order {\it divergent} Feynman diagram
which contributes to charge renormalization, we can at least adduce some%
thing of a plausibility argument.  That diagram describes the photon's
dissociation into and recombination from a virtual electron-positron pair.
We identify two successive subdiagrams of this diagram, which may be inter%
preted as the four-dimenensional Fourier transforms of, respectively, the
stress-energy contribution to the electromagnetic field from the above-
mentioned photon dissociation-recombination to a virtual pair, and the
stress-energy of the virtual pair itself.  Both of these subdiagrams carry
the logarithmic divergence of the parent diagram, but the convolution char-
acter of the second one permits calculation of its inverse Fourier transform
to obtain the above-mentioned virtual pair stress-energy---which must be high%
ly localized in space-time and have a nonintegrable singularity that produces
the divergence in its Fourier transform.  Self-gravitational effects should,
however, modify this nonintegrable $G = 0$ virtual pair stress-energy to make
it everywhere integrable, as occurs with the electromagnetic field energy
density of the electron.  Thus, with $G$ ``switched on'', its Fourier trans%
form will become convergent, and with it all the above-mentioned diagrams
for photon dissociation-recombination to a virtual pair, as well as their
contribution to charge renormalization.

We proceed now to blithely speculate that taking proper account of gravita%
tion replaces all the unresolved ultraviolet divergences of quantum elec%
trodynamics with finite $G$-dependent quantities (which, of course, must
diverge in the limit that $G\rightarrow 0$, and thus {\it cannot\/} be cal%
culated in the context of a perturbative gravitational development in non%
negative powers of $G$, such as is normally used in quantized gravitational
theory).  Indeed, we extend this speculation to all of quantum field theory,
and ask what its implications are vis-\`a-vis the conventional insistence that
any acceptable quantum field theory needs to be renormalizable.  We conclude
that under such circumstances the renormalizability requirement becomes a
{\it physically\/} sensible constraint, rather than a calculational {\it nec%
essity\/}, for narrowing the field of candidate quantum field theories which
one might try to postulate for almost any force---with one transcendent ex%
ception: there is no longer any compelling argument as to why quantized grav%
itation theory {\it itself\/} must be renormalizable.  This would be a del%
ightful d\'enouement indeed, as it has long been a bewilderment that quantized
general relativity alone, among the quantum field theories which one is com%
pelled to take seriously, turns out to be nonrenormalizable and thus osten%
sibly calculationally senseless.
\medskip
\centerline{{\bf II.}\enspace{\ssbf Self-Gravitational Corrections to Static,
Spherically Symmetric, Nonnegative}}
\centerline{\ssbf Localized Energy Densities Obtained from G = 0 Theories}
\smallskip
{}From symmetry considerations it can be shown that the general static, spher%
ically symmetric metric has a proper time interval of the form [4],
$$
d\tau^2 = B(r)\,dt^2-A(r)\,dr^2-C(r)r^2(d\theta^2+\sin^2\theta\>d\phi^2).%
\eqno{(1)}
$$
Because general relativity is a gauge theory in which the metric plays the
role of a tensorial gauge potential, the metric is not unique.  This gauge
freedom turns out to permit one to fix the metric function $C(r)$ above at
its flat-space value, $C(r) = 1$, resulting in the static, spherically sym%
metric metric of ``standard form'' (``standard gauge'' would seem to be a
more appropriate terminology),
$$
g_{\mu\nu}\,dx^{\mu}\,dx^{\nu} = d\tau^2%
= B(r)\,dt^2-A(r)\,dr^2-r^2(d\theta^2+\sin^2\theta\>d\phi^2).\eqno{(2)}
$$
In our problem the {\it source\/} of this static, spherically symmetric
curvature (i.e., the departure of the metric functions $B(r)$ and $A(r)$
from their flat-space value of unity) is, in the context of an idealized
flat-space ($G = 0$) calculation, a static, spherically symmetric, non%
negative localized energy density $\veps (r)$.  In the $G = 0$ Minkowskian
``world'', the corresponding complete stress-energy tensor may be formally
written as,
$$
T^{\mu\nu} = \veps U^\mu U^\nu,\eqno{(3)}
$$
where $U^\mu$ is the velocity four-vector, which satisfies,
$$
\eta_{\mu\nu}U^\mu U^\nu = 1,\eqno{(4)}
$$
where $\eta_{\mu\nu}$ is the Minkowski (flat-space) metric.  In our par%
ticular frame of reference, where $T^{\mu\nu}$ reduces to just its $T^{00}$
component, which is the static energy density $\veps (r)$, we have the spat%
ial components $U^i$ ($i=1,2,3$) of $U^\mu$ vanishing and $U^0 = 1$.  It is
easily verified that $T^{\mu\nu}$ meets all the flat-space criteria for a
stress-energy tensor---namely Lorentz covariance, symmetry in its indices,
and divergencelessness in the ordinary (Minkowskian) sense.  We note in
passing that the radius argument $r$ of $\veps (r)$ has the Lorentz-invar%
iant representation, $r = ((\eta_{\mu\nu}U^\mu x^\nu)^2-\eta_{\alpha\beta}%
x^\alpha x^\beta)^{1/2}$.

In the general-relativistic case, it would seem natural to try to keep
Eq.\ (3) as the stress-energy tensor which serves as the ``source'' term
in the Einstein equation for the metric (i.e., the functions $A(r)$ and
$B(r)$).  Of course, the four-velocity $U^\mu$ must then satisfy the fully
covariant generalization of Eq.\ (4),
$$
g_{\mu\nu}U^\mu U^\nu = 1,\eqno{(5)}
$$
which, for our static reference frame, now implies that $U^t = (B(r))^{-1/2}$
instead of unity, while the spatial components $U^r$, $U^\theta$, and
$U^\phi$ of course still vanish.  However, a problem now arises.  While
this generalization of Eq.\ (3) continues to be divergenceless in the ordin%
ary (Minkowskian) sense, it fails to be divergenceless in the general covar%
iant sense with the metric of Eq.\ (2).  Indeed, it turns out that,
$$
T^{\mu r}{}_{;\mu} = {1\over 2}{B'(r) \ovwp B(r)}%
{\veps (r) \ovwp A(r)},
$$
which fails to vanish unless $B'(r) = 0$ (as is the case for flat space)
or $\veps (r)$, the $G = 0$ energy density itself, vanishes.  We thus can%
not use the $T^{\mu\nu}$ of Eq.\ (3) in the Einstein equation, which is a
gauge equation that is inconsistent unless $T^{\mu\nu}{}_{;\mu}$ vanishes.
To obtain a $T^{\mu\nu}$ which is divergenceless in the generally covariant
sense, we need to generalize Eq.\ (3).  The other symmetric tensor besides
$U^\mu U^\nu$ which is natural to the problem is the metric tensor
$g^{\mu\nu}$ itself.  So we follow the precedent of static, spherically
symmetric stellar structure theory [2] and introduce a static pressure
term into the $T^{\mu\nu}$ of Eq.\ (3),
$$
T^{\mu\nu} = (\veps + p) U^\mu U^\nu - p g^{\mu\nu}.\eqno{(6)}
$$
Note that the pressure $p$ is introduced in such a way that we still have
$T^{00} = (\veps (r)/ B(r))$ in our static reference frame, exactly the
same as was the case for Eq.\ (3).  For our purposes the static pressure
$p\,(r)$ is meant to have no existence independent of that of $\veps (r)$
and nonflatness of the metric---it is introduced {\it only\/} to ensure
that $T^{\mu\nu}$ is divergenceless in the generally covariant sense, and
we thus shall require it to vanish if $\veps (r)$ does, or if the metric
functions $B(r)$ and $A(r)$ go to unity.  As $p\,(r)$ is purely a creature
of the metric and $\veps (r)$, we shall require that it be localized, as
$\veps (r)$ is.  With $p\,(r)$ in place, the offending (radial) component of
the generally covariant divergence of $T^{\mu\nu}$ becomes,
$$
T^{\mu r}{}_{;\mu} = {1 \ovwp A}\left ( p' + {1 \over 2}{B' \ovwp B}p + %
{1 \over 2}{B' \ovwp B}\veps\right ).
$$
This can be made to vanish if $p\,(r)$ is required to satisfy the linear
inhomogeneous first-order differential equation,
$$
p' + {1 \over 2}{B' \ovwp B}p = -\,{1 \over 2}{B' \ovwp B}\veps,\eqno{(7)}
$$
which may readily be reduced to quadrature after multiplying through by
the integrating factor $B^{1 \over 2}$,
$$
p\,(r) = {1 \ovwp 2B^{1 \over 2}(r)}\int_r^{r_0} d\rho\,%
{B'(\rho)\,\veps(\rho)%
\ovwp B^{1 \over 2}(\rho)}.
$$
We enforce the requirement that $p\,(r)$ be localized, as $\veps (r)$ is,
by choosing the integration constant $r_0$ to be $\infty$.  The result%
ing $p\,(r)$,
$$
p\,(r) = {1 \ovwp 2B^{1 \over 2}(r)}\int_r^{\infty} d\rho\,%
{B'(\rho)\,\veps(\rho)%
\ovwp B^{1 \over 2}(\rho)},\eqno{(8)}
$$
is localized and vanishes if $\veps (r)$ does or the space becomes flat, as
we have required.  The $T^{\mu\nu}$ of Eqs.\ (6) and (8) is now divergence%
less in the generally covariant sense and, of course, has all of its Car%
tesian components localized in our static reference frame.  It vanishes
identically when $\veps$ is put to zero, and, in the flat-space limit in
our static reference frame, satisfies $T^{00} = \veps$, with all other
components zero.

Putting this satisfactory $T^{\mu\nu}$ of Eqs.\ (6) and (8) into the Einstein
equation together with the metric form of Eq.\ (2), results, in our static
reference frame, in the following three equations for the metric functions
$B(r)$ and $A(r)$ [2],
$$
\eqalignno{
&{1\over r}{A'\ovwp A}+{1\over r^2}(A-1)=8\pi GA\veps ,&(9a)\cr
&{1\over r^2}(1-A)+{1\over r}{B'\ovwp B}=8\pi GAp,&(9b)\cr
&{1\over 2}{B^{\prime\prime}\ovwp B}-%
{1\over 4}\left (B'\over B\right )^2-{1\over 4}{B'\ovwp B}%
{A'\ovwp A}+{1\over {2r}}\left ({B'\over B}-{A'\over A}\right )=8\pi GAp,%
&(9c)\cr}
$$
where $p$ is, of course, given by Eq.\ (8).  If we multiply Eq.\ (9a)
through by $(r^2/A)$, we find that we can re-express it as,
$$
\left ( r {\left ( 1-{1 \over A} \right )} \right )'=8\pi Gr^2\veps,
$$
which can be solved for $A$,
$$
A(r)=\left ( 1-{8\pi G\over r}\int_{a_0}^r \rho^2\,d\rho\,\veps (\rho)%
\right )^{-1}.\eqno{(10)}
$$
Since $\veps (r)$ is a localized energy density, we can see from Eq.\ (10)
that $A(r)\rightarrow 1$ as $r\rightarrow \infty$ regardless of the value
of the integration constant $a_0$.  This physically proper behavior of
$A(r)$, tending toward the flat-space value of unity at large distances
from the localized energy density $\veps$, leaves us without a straight%
forward criterion at this point to pin down the value of the integration
constant $a_0$.  We shall return to the issue of determining $a_0$ be%
low---indeed it will turn out that the proper determination of $a_0$, to%
gether with the consequences which flow from that determination, comprise
the main result of this section.

First, however, we show that Eqs.\ (9a), (9b), and (9c) contain redundan%
cy---as well they ought, since there are only two unknown functions, $B(r)$
and $A(r)$, with which to solve these three equations.  If we multiply
Eq.\ (9b) through by $(B^{1 \over 2}/A)$, differentiate both sides with res%
pect to r, and then multiply the result by $(A/B^{1 \over 2})$, we obtain,
$$
\eqalignno{
&{1\ovwp 2r^2}{B'\ovwp B}(1-A)-{2\ovwp r^3}(1-A)-{1\ovwp r^2}%
\left ({A'\over A}+{B'\over B}\right )-{1\over r}{A'\ovwp A}{B'\ovwp B}\cr
&+{1\over r}\left ({B^{\prime\prime}\over B}-%
{1\over 2}{\left (B'\over B\right )^2}\right )%
=8\pi GA\left ( {1 \over 2}{B'\ovwp B}p+p'\right ).&(11)\cr}
$$
Using Eq.\ (7), we see that we can replace the right hand side of Eq.\ (11)
by $-4\pi GA\veps (B'/B)$.  Having done this, we multiply both sides of
Eq.\ (9a) by $(B'/(2B))$ and add this to the just described modified form
of Eq.\ (11) to obtain,
$$
-{2\ovwp r^3}(1-A)-{1\ovwp r^2}%
\left ({A'\over A}+{B'\over B}\right )-{1\over 2r}{A'\ovwp A}{B'\ovwp B}+%
{1\over r}\left ({B^{\prime\prime}\over B}-%
{1\over 2}{\left (B'\over B\right )^2}\right )%
=0.\eqno{(12)}
$$
Now we multiply Eq.\ (12) by $(r/2)$ and rearrange the terms to obtain,
$$
{1\over 2}{B^{\prime\prime}\ovwp B}-{1\over 4}{\left (B'\over B\right )^2}%
-{1\over 4}{A'\ovwp A}{B'\ovwp B}%
-{1\over 2r}\left ({A'\over A}+{B'\over B}\right )%
-{1\ovwp r^2}(1-A)=0.\eqno{(13)}
$$
We have derived Eq.\ (13) from Eqs.\ (9b) and (9a).  However, Eq.\ (13) fol%
lows as well from Eqs.\ (9b) and (9c)---these have identical right hand
sides, so we may equate their left hand sides, which also yields Eq.\ (13).

We have thus demonstrated that only {\it two\/} of the three equations
(9a), (9b), and (9c) are independent.  (As we mentioned above, this is
fortunate, as we deal with only {\it two\/} metric functions $B(r)$ and
$A(r)$.)  In practice we shall be using Eq.\ (13) above and Eq.\ (9a)
(whose solution is given by Eq.\ (10)) as the two equations which deter%
mine $B$ and $A$.  The redundancy we have demonstrated is a consequence
of the gauge ambiguity of the Einstein equation---this is what permits
us to set the third metric function $C(r)$ to unity in the general static,
spherically symmetric metric given by Eq.\ (1), producing the ``standard
form'' (or ``standard gauge'') given by Eq.\ (2).

It is interesting to note that if we multiply Eq.\ (13) through by
$B^{1\over 2}$, it can be reexpressed as a homogeneous {\it linear\/}
second-order equation for $B^{1\over 2}$,
$$
\left (B^{1\over 2}\right )^{\prime\prime}-%
\left ({1\over r}+{1\over 2}{A' \ovwp A}\right )%
\left (B^{1\over 2}\right )^{\prime}-%
\left ({1\ovwp r^2}(1-A)+{1\over 2r}{A' \ovwp A}\right )B^{1\over 2}=%
0.\eqno{(14)}
$$
Since the energy density $\veps (r)$ and stress-energy tensor $T^{\mu\nu}$
are localized, we expect the gravitational forces to be weak at large $r$,
and for $B^{1\over 2}$ to asymptotically approach a Newtonian limiting
form [5],
$$
B^{1\over 2}\sim 1-{G{\cal M}\over r}%
\quad \hbox{\rm as $r\to\infty$},\eqno{(15)}
$$
where ${\cal M}$ is the effective total gravitating mass of the system.
Also at large $r$ we have from Eq.\ (10) that the asymptotic form of $A$ is,
$$
A(r)\sim 1+{8\pi G\over r}\int_{a_0}^\infty \rho^2\,d\rho\,\veps (\rho)%
\quad \hbox{\rm as $r\to\infty$},\eqno{(16)}
$$
If we multiply Eq.\ (14) through by $r^2$, and insert into the resulting
equation the large $r$ asymptotic forms of Eqs.\ (15) and (16), the terms
through first order in $r^{-1}$ yield,
$$
{\cal M}=4\pi\int_{a_0}^\infty r^2\,dr\,\veps (r).\eqno{(17)}
$$
We thus see that the total gravitating mass ${\cal M}$ of our system depends
on the constant of integration $a_0$.  In the case where $\veps (r)$ is
{\it everywhere\/} sufficiently small that the whole problem may be treated
throughout in the weak-gravitational Newtonian limit, we know from Newton%
ian gravity theory that the total gravitating mass comes out to be simply
the $G = 0$ energy density $\veps (r)$ integrated over all space,
$$
{\cal M}=\int d^3\vec r\,\veps (r)=4\pi\int_0^\infty r^2\,dr\,\veps (r).
$$
Thus, in the weak-gravitational limit, we may conclude that $a_0=0$.
Can we now conclude that $a_0$ is {\it always} equal to zero?  We can
see from Eq.\ (10) that if $\veps (r)$ is not everywhere sufficiently
small, it is possible for the metric function $A$ to develop singularities.
They occur at values of $r$ where
$$
{8\pi G\over r}\int_{a_0}^r \rho^2\,d\rho\,\veps (\rho)=1\eqno{(18)}
$$
is satisfied.  Since $\veps (r)$ is localized and nonnegative, choosing
$a_0$ sufficiently {\it large\/} ensures that there exists no $r$ where
the singularity condition (18) can be satisfied.  So if we want to suppress
singularities in the metric function $A$, it is not always possible to
stick with the weak-gravitational prescription that $a_0$ be set to zero.
Of course, because of the gauge ambiguity, metric functions such as $A$
are not unique, which makes their physical meaning somewhat obscure.  Be%
fore we try to draw definite conclusions on the matter of how $a_0$ is to
be chosen, we first need to see if there are quantities having clearcut
physical meaning which as well become singular when Eq.\ (18) is satisfied.
An ideal such quantity would be an energy density which can be interpret%
ed as properly incorporating the self-gravitational corrections to our
$G = 0$ energy density $\veps (r)$.  Because $\veps (r)$ and our source
stress-energy tensor $T^{\mu\nu}$ are localized, it happens to indeed be
possible to construct such a full energy density.

When the Einstein equation's source stress-energy tensor $T^{\mu\nu}$ is
localized, and the metric tensor is constrained to approach the Minkow%
skian flat-space value at large distances from that source (the situation
for our problem), it is legitimate to view that equation as describing the
dynamics of an ordinary classical field (a Minkowskian tensor one, which is
the metric minus the Minkowski flat-space metric) in the context of Minkow%
skian flat space.  The dynamics of this field are, to be sure, highly non%
linear, and have a gauge character, but the well-known classical non-Abelian
gauge theories share these particular features as well, and they are always
viewed as playing out their dynamics in the context of Minkowskian flat
space.  Field theories in flat space generally possess a total stress-energy
tensor which is divergenceless in the ordinary (Minkowskian) sense, a proper%
ty which allows the normal definition of conserved total energy, momentum,
and angular momentum.  This turns out to be the case as well for asymptotic%
ally flat gravity theory, where this ordinary total stress-energy tensor
$\tau^{\mu\nu}$ is constructed from the linear, purely second-order deriva%
tive part of the Cartesian components of the Einstein tensor [6], an object
which is divergenceless in the ordinary, Minkowskian, sense (the full non%
linear Einstein tensor is, of course, only divergenceless in the generally
covariant sense, which does not produce a conserved total energy, momentum,
and angular momentum).  We are particularly interested in the ordinary total
energy density $\tau^{00}$ (which is the net physical energy density result%
ing from our $G = 0$ input energy density $\veps$ plus that from its self-%
gravitational interaction), since the total gravitating mass ${\cal M}$ must
be the total localized physical energy, i.e., the integral over all space of
$\tau^{00}$.  This, together with Eq.\ (17), implies that,
$$
\int d^3\vec r\,\tau^{00}=4\pi\int_{a_0}^\infty r^2\,dr\,\veps (r).\eqno{(19)}
$$
Since $\tau^{00}$ is constructed from part of the Einstein tensor (the linear
part), and thus from the metric, it will be interesting to see whether it
shares the singularities which occur in our metric function $A$ when Eq.\ (18)
is satisfied.  Such singularities in $\tau^{00}$, if they are nonintegrable,
could disrupt the physically required relationship given by Eq.\ (19), and
thus could constrain $a_0$ to values sufficiently large so that the singu%
larity condition of Eq.\ (18) cannot be satisfied.  However, we are getting
ahead of ourselves: we now need to calculate $\tau^{\mu\nu}$ and then inves%
tigate the properties of $\tau^{00}$.

In the flat-space approach to gravity theory, the dynamic gravitational
gauge field $h_{\mu\nu}$ is given by the metric tensor $g_{\mu\nu}$ in
Cartesian coordinates, minus the flat-space (Minkowski) metric
$\eta_{\mu\nu}$,
$$
h_{\mu\nu}\equiv g_{\mu\nu}-\eta_{\mu\nu}.
$$
This flat-space approach only turns out to be fully viable in situations
where $h_{\mu\nu}$ is localized [6], as is the case for our problem.
Following reference [6], we exhibit the part of the Ricci curvature tensor
which is linear in $h_{\mu\nu}$,
$$
R^{(1)}{}_{\mu\kappa}={1\over 2}\left (%
{\partial^2 h^\lambda{}_\lambda\over\partial x^\mu\partial x^\kappa}%
-{\partial^2 h^\lambda{}_\mu\over\partial x^\lambda\partial x^\kappa}%
-{\partial^2 h^\lambda{}_\kappa\over\partial x^\lambda\partial x^\mu}%
+{\partial^2 h_{\mu\kappa}\over\partial x^\lambda\partial x_\lambda}%
\right ),
$$
where we adopt the useful convention that indices on $h_{\mu\nu}$,
$\partial /\partial x^\lambda$, and $R^{(1)}{}_{\mu\kappa}$ are raised
and lowered with the Minkowski metric $\eta_{\mu\nu}$ rather than
$g_{\mu\nu}$.

Using $R^{(1)}{}_{\mu\kappa}$, we obtain $G^{(1)}{}_{\mu\kappa}$, the part
of the Einstein tensor which is linear in $h_{\mu\nu}$ (and, like $R^{(1)}%
{}_{\mu\kappa}$, purely second-order in the derivatives of $h_{\mu\nu}$),
$$
G^{(1)}{}_{\mu\kappa}=R^{(1)}{}_{\mu\kappa}-%
{1\over 2}\eta_{\mu\kappa}R^{(1)}{}^\lambda{}_\lambda.
$$
It can be verified that $G^{(1)}{}_{\mu\kappa}$ is symmetric in its indices
and divergenceless in the ordinary (Minkowskian) sense.  Thus we adopt as
the definition of $\tau^{\mu\nu}$, the ordinary total stress-energy tensor
[6],
$$
\tau^{\mu\nu}\equiv -{1\over 8\pi G}\eta^{\mu\alpha}\eta^{\nu\beta}%
G^{(1)}{}_{\alpha\beta}.
$$
If $h_{\mu\nu}$ is localized, this divergenceless (in the Minkowskian
sense) $\tau^{\mu\nu}$ is localized as well, and permits construction
in the usual flat-space way of conserved total energy, momentum, and
angular momentum.  For our purposes, the physical quantity of command%
ing interest is the total energy, which is obtained by integrating the
localized $\tau^{00}$ over all space, as in Eq.\ (19).

For our problem, with the metric of Eq.\ (2), $h_{\mu\nu}$ is given by,
$$
\eqalignno{
h_{00}&=B(r)-1,&(20a)\cr
h_{ij}&=(1-A(r)){x^i x^j\ovwp r^2},&(20b)\cr
h_{i0}&=h_{0i}=0\quad\hbox{\rm where $i,j=1,2,3$ in (b) \& (c)}.&(20c)\cr}
$$
As Eqs.\ (15) and (16) imply that $B(r)\to 1$ and $A(r)\to 1$ as $r\to\infty$,
our $h_{\mu\nu}$ is indeed localized, and the resulting ordinary total
stress-energy tensor $\tau^{\mu\nu}$ will be fully physically meaningful in
the familiar flat-space sense [6].

Proceeding now to apply the above-given formulas for calculating $\tau^%
{\mu\nu}$ from $h_{\mu\nu}$, we obtain from Eq.\ (20), after many tedious
differentiations and index contractions,
$$
\eqalignno{
\tau^{00}&={1\ovwp 8\pi G}\left [{A'\over r}+{1\over r^2}(A-1)\right ]=%
{1\ovwp 8\pi Gr^2}(r(A-1))',&(21a)\cr
\tau^{ij}&={1\ovwp 16\pi G}\left [%
\left (B^{\prime\prime}+{B'\over r}-{A'\over r}\right )\delta_{ij}-%
\left (B^{\prime\prime}-{B'\over r}-{A'\over r}+{2\over r^2}(A-1)\right )%
{x^i x^j\ovwp r^2}\right ],&(21b)\cr
\tau^{i0}&=\tau^{0i}=0\quad\hbox{\rm where $i,j=1,2,3$ in (b) \& (c)}.%
&(21c)\cr}
$$
The static stress tensor $\tau^{ij},\; i,j=1,2,3,$ is symmetric in its two
indices and identically divergenceless in three-dimensional Euclidean space,
as indeed it must be.  The three-momentum density $\tau^{i0},\; i=1,2,3,$
vanishes, as it physically ought to in such a static situation.  The energy
density $\tau^{00}$ of Eq.\ (21a) turns out to depend {\it only\/} on the
metric function $A(r)$, which we know explicitly from Eq.\ (10).  Thus we can
write,
$$
\tau^{00}(r)={1\over r^2}\left (\int_{a_0}^r \rho^2\,d\rho\,\veps (\rho)%
\over 1-{8\pi G\over r}%
\int_{a_0}^r \rho^2\,d\rho\,\veps (\rho)\right )',\eqno{(22a)}
$$
or, explicitly,
$$
\tau^{00}(r)={\veps (r)-8\pi G\left ({1\over r^2}%
\int_{a_0}^r \rho^2\,d\rho\,\veps (\rho)\right )^2%
\over \left (1-{8\pi G\over r}%
\int_{a_0}^r \rho^2\,d\rho\,\veps (\rho)\right )^2}.\eqno{(22b)}
$$
It is interesting to note, from Eq.\ (22b), that if $a_0>0$, then any
nonnegative $G = 0$ energy density $\veps (r)$ which is {\it nonintegra%
bly singular\/} as $r\to 0$ results in the universal asymptotic behavior
$\tau^{00}\sim -(8\pi Gr^2)^{-1}$ as $r\to 0$, which {\it is\/} integrable
as $r\to 0$, and {\it negative\/} there to boot!  We thus see that self-%
gravitational effects are indeed in principle capable of overwhelming intrac%
table positive singularities of $G = 0$ energy densities with negative
gravitational energy that renders such singularities harmless.

There still remains the question of whether strong self-gravitational
effects compel a non-Newtonian positive choice for $a_0$.  Examination
of Eq.\ (22a) shows that $\tau^{00}(r)$ is nonintegrably singular at
{\it positive\/} values of $r$ where Eq.\ (18) is satisfied, just as the
metric function $A(r)$ is singular at such positive values of $r$.  As
we have already stated, the fact that $\veps (r)$ is localized and non%
negative ensures that for sufficiently large $a_0$ there exists no $r$
which satisfies the singularity condition of Eq.\ (18).  Thus to keep
$\tau^{00}$ integrable, we can indeed sometimes be compelled to choose
$a_0>0$.

Provided that we have now chosen $a_0$ so as to render $\tau^{00}(r)$
free from nonintegrable singularities, we may use Eq.\ (22a) to evalu%
ate the total energy (mass) of the system,
$$
\eqalign{
\int d^3\vec r\,\tau^{00}&=4\pi\int_0^\infty r^2\,dr\,{1\over r^2}%
\left (\int_{a_0}^r \rho^2\,d\rho\,\veps (\rho)%
\over 1-{8\pi G\over r}%
\int_{a_0}^r \rho^2\,d\rho\,\veps (\rho)\right )'\cr
&=\left .4\pi\int_{a_0}^r \rho^2\,d\rho\,\veps (\rho)%
\over 1-{8\pi G\over r}%
\int_{a_0}^r \rho^2\,d\rho\,\veps (\rho)\right\arrowvert_{r\to\infty}
-\left .4\pi\int_{a_0}^r \rho^2\,d\rho\,\veps (\rho)%
\over 1-{8\pi G\over r}%
\int_{a_0}^r \rho^2\,d\rho\,\veps (\rho)\right\arrowvert_{r\to 0}\cr
&=4\pi\int_{a_0}^\infty\rho^2\,d\rho\,\veps (\rho)\cr}
$$
Thus we have indeed arrived at Eq.\ (19) (for which we had previously
given a nonformal, physically motivated argument), but with the pre%
viously anticipated caveat that $a_0$ must have been chosen sufficiently
large that the $\tau^{00}$ nonintegrable singularity condition given by
Eq.\ (18) has no solutions.  The bottom-line question now presents itself:
how is $a_0$ to be {\it uniquely\/} determined in {\it all\/} situations?
The two constraints which we have so far obtained on the choice of $a_0$
is that it must be zero (its smallest possible value, since, of course,
$r\ge 0$) in the weak-gravity Newtonian limit, and large enough to avoid
the occurrence of nonintegrable singularities in $\tau^{00}(r)$ when
gravitational effects become strong.  We may obtain yet another constraint
on the choice of $a_0$ from the general principle of {\it continuity\/} in
classical physics.  Thus, we expect $a_0$ (and, perhaps more to the point,
the total mass ${\cal M}$ of the system, which depends on $a_0$, as Eq.\ (17)
demonstrates) to vary in a continuous way if $\veps (r)$ is varied continu%
ously.  If $\veps (r)$ is sufficiently small everywhere, we will be in the
weak-gravitational Newtonian regime, where $a_0$ must be zero.  Now, as we
continuously vary $\veps (r)$ to take on larger and larger values, we must
eventually reach a point where the choice $a_0=0$ results in $\tau^{00}(r)$
developing a nonintegrable singularity, which then forces us to make $a_0$
sufficiently positive.  But just how positive?  The only choice which can
be consistent with the continuity requirement is that $a_0$ must be the
{\it smallest\/} positive value for which the singularity condition of
Eq.\ (18) has no solution.  Only such a minimum recipe permits $a_0$ to fall
{\it continuously\/} from positive values to the required Newtonian limit
value of zero as $\veps (r)$ is continuously weakened.  We may thus write
down a technical definition of $a_0$ as follows,
$$
a_0(\epsilon)=\min\left\{\,a\mid a\ge 0\;{\rm and}\;\max_{0\le r<\infty}%
\left ({8\pi G\over r}\int_a^r\rho^2\,d\rho\,\veps (\rho)\right )%
\le 1-\epsilon\,\right\},\eqno{(23a)}
$$
and
$$
a_0=\lim_{\epsilon\to 0+}\,a_0(\epsilon).\eqno{(23b)}
$$
For every $a_0(\epsilon)$ where $\epsilon > 0$, $\tau^{00}(r)$ has no
nonintegrable singularities.  For sufficiently strong $\veps (r)$, as
$\epsilon$ gets smaller and smaller the singularity condition
of Eq.\ (18) will come closer and closer to being
satisfied.  The radius $r = r_s$ for which Eq.\ (18) tends toward being
satisfied as $\epsilon\to 0+$ in Eqs.\ (23), can be regarded as the loca%
tion of an ``incipient'' singularity of $\tau^{00}(r)$, and interpreted
as the Schwarzschild radius of a black hole.  In particular, we can re%
write Eq.\ (17) for the total mass ${\cal M}$ of the system to refer only
to this critical radius $r_s$ instead of to $a_0$ itself.  First we split
up the integration interval of Eq.\ (17) as follows,
$$
{\cal M}=4\pi\int_{a_0}^{r_s}r^2\,dr\,\veps(r)+%
4\pi\int_{r_s}^\infty r^2\,dr\,\veps(r).
$$
Since Eq.\ (18) is (``incipiently'') satisfied at $r = r_s$, we have that,
$$
{8\pi G\over r_s}\int_{a_0}^{r_s}r^2\,dr\,\veps(r)=1,\eqno{(24)}
$$
and thus,
$$
{\cal M}={r_s\over 2G}+%
4\pi\int_{r_s}^\infty r^2\,dr\,\veps(r).\eqno{(25)}
$$
We may give a loose intuitive interpretation of Eq.\ (25) by saying that the
total mass of the system comes from the $G = 0$ contribution for the region
outside the critical radius $r_s$, plus the standard expression for the
mass of a black hole of Schwarzschild radius $r_s$ (the mass of a
black hole equals its Schwarzschild radius divided by $(2G)$).  In the
weak-gravitational Newtonian limit, of course $a_0 = 0$, Eq.\ (18) is not
satisfied (even ``incipiently'') for any $r\ge 0$, and the critical radius
$r_s$ is properly taken to be zero in Eq.\ (25), making it consistent
with Eq.\ (17) in this weak-gravitational case.

In many situations it is possible to obtain the critical radius $r_s$
directly from $\veps (r)$, without reference to $a_0$.  To demonstrate
this we define, because of the content of Eqs.\ (23),
$$
f(r;a)\equiv{8\pi G\over r}\int_a^r\rho^2\,d\rho\,\veps(\rho).\eqno{(26)}
$$
It will frequently be the case that the maximum in $r$ of $f(r;a)$, which
is required in Eq.\ (23a), occurs where the derivative with respect to $r$
of $f(r;a)$ vanishes.  The radius $r_{\max}$, where the maximum of $f(r;a)$
occurs, will in that case satisfy,
$$
8\pi Gr_{\max}\veps(r_{\max})={1\ovwp r_{\max}}f(r_{\max};a).\eqno{(27)}
$$
In general, $r_{\max}$ is a function of $a$, and may be written $r_{\max}(a)$.
When $a=a_0$, $r_{\max}(a=a_0)=r_s$ and, from Eqs.\ (26) and (24),
$$
\left .f(r_{\max};a)\right\arrowvert_{a=a_0}=f(r_s;a_0)=1.\eqno{(28)}
$$
Thus, putting $a$ to $a_0$ in Eq.\ (27) yields,
$$
8\pi Gr_s\veps(r_s)={1\over r_s},
$$
which implies that,
$$
4\pi r_s\!\!\!{}^2\>\veps(r_s)=(2G)^{-1}.\eqno{(29)}
$$
Eq.\ (29) has given us the condition for the critical radius $r_s$ that
was mentioned in Section I, namely that it occurs where the $G = 0$
energy per radial distance attains the limiting value $(2G)^{-1}$.
Eq.\ (25) may then be further loosely interpreted as saying that the
energy per radial distance in effect ``freezes'' at this $(2G)^{-1}$
limiting value within the black hole of radius $r_s$, while it takes
on its $G = 0$ value outside the black hole---up to the point that it
equals $(2G)^{-1}$, which defines the black-hole boundary radius $r_s$.
Of course there exists no black-hole interior region to be concerned
with if the $G = 0$ energy per radial distance is always less than the
limiting value $(2G)^{-1}$.  In that case, Eqs.\ (23) imply that $a_0 = 0$
(of course $r_s = 0$ as well for consistency between Eqs.\ (25) and (17)),
so the total mass ${\cal M}$ of the system is equal to its $G = 0$ value,
just as in the Newtonian limit.

Eqs.\ (29) and (25) will be all that we require for our discussion of the
electromagnetic field contribution to the electron's mass in the next sec%
tion.  It does seem disappointing that such very simple and intuitively
appealing results have entailed such a lengthy, ponderous, and sometimes
subtle derivation.

Before we go on to the discussion of the electromagnetic field contribution
to the electron mass, it is interesting to ask how our stress-energy tensor
$\tau^{\mu\nu}$ looks in the weak-gravitational Newtonian limit.  Of course
we know that we must take $a_0 = 0$ in that situation.  We also need to
approximate the metric functions $B(r)$ and $A(r)$ through just first order
in $G$.  In the Newtonian limit, we further know that [5],
$$
B(r)=1+2\phi(r),\eqno{(30)}
$$
where
$$
\nabla^2\phi=\phi^{\prime\prime}+{2\over r}\phi'=%
{1\over r^2}\left (r^2\phi'\right )'=4\pi G\veps(r).\eqno{(31)}
$$
Thus,
$$
r^2\phi'(r)=4\pi G\int_0^r\rho^2\,d\rho\,\veps(\rho).
$$
To first order in $G$ with $a_0 = 0$ (Newtonian limit), Eq.\ (10) becomes,
$$
A(r)=1+{8\pi G\over r}\int_0^r\rho^2\,d\rho\,\veps(\rho),
$$
which, together with the previous equation, yields,
$$
A(r)=1+2r\phi'(r).\eqno{(32)}
$$
If we put the Newtonian limiting forms for $B(r)$ and $A(r)$ given by
Eqs.\ (30) and (32) into Eqs.\ (21) for $\tau^{\mu\nu}$, and bear in mind
that the Newtonian gravitational potential $\phi$ satisfies Eq.\ (31),
we obtain,
$$
\tau^{00}=\veps(r)
$$
and
$$
\tau^{ij}=0\quad\hbox{\rm for $i,j=1,2,3$}.
$$
Thus the Newtonian limit of $\tau^{\mu\nu}$ reduces to its $G = 0$ case.
We see that spherically symmetric Newtonian gravity theory entails no
self-gravitational correction effects whatsoever.
\medskip
\centerline{{\bf III.}\enspace{\ssbf The Electromagnetic Field Contribution
to the Electron Mass}}
\smallskip
The classical point-charge model for the electron has the familiar electro%
static field,
$$
\vec E(\vec r)=-{e\vec r\over |\vec r|^3},
$$
which produces the static energy density,
$$
\veps(r)={|\vec E(\vec r)|^2\over 8\pi}={e^2\over 8\pi r^4}.\eqno{(33)}
$$
Eq.\ (29) for the critical radius $r_s$ thus reads,
$$
{e^2\over 2r_s\!\!\!{}^2}={1\over 2G}
$$
or
$$
r_s=\left (Ge^2\right )^{1\over 2}\approx 1.4\times 10^{-21}\>{\rm fm}.%
\eqno{(34)}
$$
Eq.\ (25) for this electrostatic energy contribution to the electron mass
then becomes,
$$
\delta m= {\left (Ge^2\right )^{1\over 2}\over 2G}+%
4\pi\int_{\left (Ge^2\right )^{1\over 2}}^\infty dr\,{e^2\over 8\pi r^2}=%
\left (e^2\over G\right )^{1\over 2},\eqno{(35)}
$$
which, unfortunately, is about $2\times 10^{21}$ times the measured elec%
tron mass!  Clearly the classical point-charge model is far out of its
depth in this issue.

In quantum electromagnetic field theory, however, charge smearing and
shielding arising from spontaneous vacuum virtual pair and photon pro%
duction reduces $\veps (r)$ well below the value given by Eq.\ (33) at
$r$ smaller than an electron Compton wavelength.  In this case, $\delta m$
(for $G = 0$) is supposed to be given by the Feynman diagram of Fig.\ 1,
with the external electron line taken to be on shell and at rest.
The calculation diverges, but only logarithmically, which is a far
weaker divergence than the linear divergence of the $G = 0$ classical
point-charge model.

Since we don't know how to do our ultrastrong, black-hole gravity
theory in the context of four-momentum space, which is the natural
environment for Feynman diagrams, we need to transform the integrand
of the $\delta m$ Feynman diagram into an energy density in configur%
ation space.  The first order of business will be to carry out the
formal integration over the $k^0$ component of the virtual photon
four-momentum in the loop part of the Feynman diagram for $\delta m$.
The result of this, of course, is a $d^3\vec k$ integral expression
for $\delta m$, whose integrand turns out to be nonnegative and
spherically symmetric (i.e., it depends only on $|\vec k|$).  We inter%
pret this integrand as an energy density in $\vec k$-space, and find that
it lends itself to transformation to a nonnegative, spherically symmetric
energy density in $\vec r$-space by the appropriate Fourier technique.
Of course both the $\vec k$-space and the $\vec r$-space energy densities
yield a logarithmically divergent $\delta m$ ($G = 0$ theory), but we
readily obtain the self-gravitational critical cutoff radius $r_s$ for
the $\vec r$-space energy density, and then proceed to calculate a finite
$\delta m$ in terms of $G$ and the bare mass and charge of the quantized
electromagnetic field theory.  We set $\hbar$ as well as $c$ to unity in
the calculations which follow.

The Feynman diagram of Fig.\ 1 for $\delta m$ is [3],
$$
\delta m= -2ie^2\int{d^3\vec k\over (2\pi)^3}%
\int_{-\infty}^\infty dk^0\,{1\ovwp k^2+i\epsilon}%
{\gamma_\nu(\slsh p-\slsh k+m)\gamma^\nu\over%
\left ((p-k)^2-m^2+i\epsilon\right )}\> ,\eqno{(36)}
$$
where the external electron of momentum p is ``on shell'' (i.e., satisfies
the Dirac equation, so that $\slsh p$ may be set to $m$) and at rest, i.e.,
$p=\left (m,\vec 0\right )$.  The upshot of both of these things together
is that $\gamma^0$ may be set to unity.  Standard Dirac matrix identities
[7] imply that,
$$
\gamma_\nu(\slsh p-\slsh k+m)\gamma^\nu= -2\slsh p + 2\slsh k + 4m.
$$
Since we can put $\slsh p$ to $m$ and $\gamma^0$ to unity, the above
may be simplified further,
$$
\gamma_\nu(\slsh p-\slsh k+m)\gamma^\nu\longrightarrow%
2m + 2k_o - 2\vec\gamma\cdot\vec k.
$$
Also, because $p=\left (m,\vec 0\right )$, $(p-k)^2-m^2+i\epsilon=%
k_o\!\!\!{}^2-2mk_o-|\vec k|^2+i\epsilon$.  Thus,
$$
\delta m= -4ie^2\int{d^3\vec k\over (2\pi)^3}%
\int_{-\infty}^\infty dk_o\,{m+k_o-\vec\gamma\cdot\vec k\over%
\left (k_o\!\!\!{}^2-|\vec k|^2+i\epsilon\right )%
\left (k_o\!\!\!{}^2-2mk_o-|\vec k|^2+i\epsilon\right )}\> .\eqno{(37)}
$$
We can clearly drop the $\vec\gamma\cdot\vec k$ term from the numerator
of the integrand of Eq.\ (37), as it has negative parity in $\vec k$ and
the integrand factors which multiply it are of positive parity in $\vec k$
(they depend only on $|\vec k|$).  The next step is to fully factorize the
denominator of the integrand,
$$
\eqalignno{
\delta m= -4ie^2\int &{d^3\vec k\over (2\pi)^3}
\int_{-\infty}^\infty dk_o\,{m+k_o\over%
\left (k_o-\left (|\vec k|-i\epsilon\right )\right )%
\left (k_o-\left ( -|\vec k|+i\epsilon\right )\right )}\;\times\cr
&{1\over\left (k_o-\left (m+\sqrt{|\vec k|^2+m^2}\>-i\epsilon\right )\right )%
\left (k_o-\left (m-\sqrt{|\vec k|^2+m^2}\>+i\epsilon\right )\right )}\> .%
&(38)\cr}
$$
The $k_o$ integration is now carried out by using the residue calculus.
If, for example, the $k_o$ contour is closed with a semicircle {\it above}
the real axis, then one calculates the residues at the two roots which
have the {\it negative} real parts, i.e., $-|\vec k|$ and
$m-\sqrt{|\vec k|^2+m^2}\,$, as these are the ones with infinitesimal
{\it positive} imaginary parts.

The upshot of the $k_o$ integration and gathering of terms turns out
to be,
$$
\delta m= 2\pi e^2 m\int{d^3\vec k\over (2\pi)^3}\,%
{f\left (|\vec k|^2/m^2\right )\over |\vec k|^2%
\left (|\vec k|^2+m^2\right )^{1\over 2}}\>,%
\eqno{(39a)}
$$
where,
$$
f(x)\equiv 1-x+\sqrt{x^2+x}.\eqno{(39b)}
$$
We note that $f(x)$ increases monotonically from unity at $x=0$ toward its
asymptotic value of $3/2$ as $x\to\infty$.  As was mentioned above, the in%
tegrand in Eq.\ (39a) is nonnegative and depends only on $|\vec k|$ (i.e.,
is spherically symmetric).  It may be interpreted as the electron's elec%
tromagnetic field energy density in $\vec k$-space.  The integral over
$\vec k$-space of a nonnegative, spherically symmetric integrand can be
reexpressed as the integral over $\vec r$-space of an integrand which is
also nonnegative and spherically symmetric.  The $\vec r$-space integrand
is obtained by squaring the Fourier transform of the square root of the
$\vec k$-space integrand.  Thus it may be shown to formally follow from
Eq.\ (39a) that,
$$
\delta m=\int d^3\vec r\;\;\veps (r),\eqno{(40a)}
$$
where,
$$
\veps (r)= 2\pi e^2 m\left (\int{d^3\vec k\over (2\pi)^3}\,%
\exp\left (i\vec k\cdot\vec r\right )%
{\left (f\left (|\vec k|^2/m^2\right )\right )^{1\over 2}\over
|\vec k|\left (|\vec k|^2+m^2\right )^{1\over 4}}%
\right )^2.\eqno{(40b)}
$$
We note here in passing that although Eq.\ (39a) for $\delta m$ is
logarithmically divergent, and Eq.\ (40a) for $\delta m$ must thus
prove to be so as well, there is no convergence problem for the
Fourier transform expression in Eq.\ (40b) for the configuration space
energy density $\veps (r)$---the object we are actually interested in.
We could have consistently inserted cutoff functions to avoid dealing
with divergent integrals, but $\veps (r)$ is clearly finite and well-def%
ined, independent of cutoff technique.

It is possible to work out the asymptotic behaviors for small and large
$r$ of the energy density $\veps (r)$ of Eq.\ (40b).  The results are,
$$
\veps (r)\sim {3\ovwp 2}{e^2 m\over 4\pi^2 r^3}\quad{\rm for}\;
r\ll {1\over m},\eqno{(41a)}
$$
and
$$
\veps (r)\sim {(2e/\pi)^2 \over 8\pi r^4}\quad{\rm for}\;
r\gg {1\over m}.\eqno{(41b)}
$$
Eq.\ (41a) shows that this quantum field theoretic energy density is less
singular as $r\to 0$ than that of the classical, point-charge model, as
given by Eq.\ (33).  Its $r^{-3}$ asymptotic behavior in this region makes
for the logarithmic divergence of $\delta m$ mentioned above.  Effective
smearing and shielding of the electron's charge due to spontaneous vacuum
production of a pair and photon, as shown in Fig.\ (2a), which is one of
the two ``time-ordered'' versions of Fig.\ (1), is responsible for this
amelioration of the $r\to\ 0$ singularity.  Eq.\ (41b) shows that at large
$r$, the behavior of $\veps (r)$ is very similar to that of the classical
point-charge model, Eq.\ (33), with the exception of an overall constant
factor which is less than unity.  This represents a finite renormalization
of the bare charge (by a factor of $(2/\pi)$, which is less than uni%
ty) that is also due to effective charge shielding by the spontaneously
produced virtual pair.

We now obtain the critical radius $r_s$ by putting the small r asymptotic
form of Eq.\ (41a) for $\veps (r)$ into Eq.\ (29),
$$
{3\ovwp 2}{e^2 m\over\pi r_s}={1\ovwp 2G},
$$
or
$$
r_s={3Ge^2 m\over\pi}.\eqno{(42)}
$$
As the expression (40b) for $\veps (r)$ is not very tractable, we make
an approximate model for $\veps (r)$ by piecing together its two asymp%
totic forms given by Eqs.\ (41a) and (41b) at the point where these cross.
This results in the continuous approximation,
$$
\veps (r)\approx\cases{3e^2m\left (8\pi^2 r^3\right )^{-1},%
&if $r\le 4 (3\pi m)^{-1}$\cr
                       e^2\left (2\pi^3 r^4\right )^{-1},
&if $r\ge 4 (3\pi m)^{-1}$.\cr}%
\eqno{(43)}
$$
Putting Eqs.\ (42) and (43) into Eq.\ (25), we arrive at,
$$
\delta m\approx 3e^2 m (2\pi)^{-1}
+\int_{3Ge^2 m \pi^{-1}}^{4 (3\pi m)^{-1}} 3e^2 m (2\pi r)^{-1}\,dr
+\int_{4(3\pi m)^{-1}}^\infty 2e^2\left (\pi^2r^2\right )^{-1}\,dr\eqno{(44a)}
$$
or
$$
\delta m\approx {3e^2 m\over 2\pi}\left (\ln{4\ovwp 9Ge^2 m^2} + 2\right ).%
\eqno{(44b)}
$$
At this stage it is convenient to define the dimensionless quantity $\Delta$,
$$
\Delta\equiv {\delta m\over m},\eqno{(45)}
$$
and rewrite Eq.\ (44b) as,
$$
\Delta\approx {3e^2\over\pi}\left (\ln{2\ovwp 3m\sqrt{Ge^2}} + 1\right )%
\eqno{(46)}
$$
The quantities $e$ and $m$ in our results of Eqs.\ (42), (44), and (46)
are the ``bare'' mass and charge parameters of the theory.  We have al%
ready seen from the comparison of the large $r$ asymptotic behavior of
$\veps (r)$ given by Eq.\ (41b) with the expected behavior given by
Eq.\ (33), that the charge actually measured at large distance, $e_m$,
will be less than the bare charge $e$ by a factor of $2/\pi$,
$$
e_m = {2e\ovwp \pi}.\eqno{(47)}
$$

Also, of course, the measured mass $m_m$ of the electron will not be
the bare mass m, but the bare mass plus the electromagnetic field
contribution $\delta m$,
$$
m_m=m+\delta m=m(1+\Delta).\eqno{(48)}
$$
In terms of the measured quantities and $\Delta$, our bare charge and
mass become,
$$
e={\pi\ovwp 2}e_m,\eqno{(49a)}
$$
$$
m={1\ovwp 1 + \Delta}m_m.\eqno{(49b)}
$$
Our electromagnetic contribution to the electron mass is likewise,
$$
\delta m=\Delta m={\Delta\ovwp 1+\Delta}m_m.\eqno{(50)}
$$
Using Eqs.\ (49) to rewrite Eqs.\ (42) and (46) in terms of the measured
charge, mass, and $\Delta$, we arrive at,
$$
r_s={3\pi Ge_m\!\!\!{}^2 m_m\over 4(1+\Delta)},\eqno{(51)}
$$
and,
$$
\Delta\approx {3\pi e_m\!\!\!{}^2\ovwp 4}\left (%
\ln{4\ovwp 3\pi m_m\sqrt{Ge_m\!\!\!{}^2}}+\ln(1+\Delta)+1\right ).\eqno{(52)}
$$
Eq.\ (52) is an implicit equation for $\Delta$, which can be solved numer%
ically by iteration.  If we put in numbers, it reads,
$$
\Delta\approx 0.931 + 0.0172\ln(1 + \Delta),\eqno{(53)}
$$
which, iterated a couple times, yields,
$$
\Delta\approx 0.942.\eqno{(54)}
$$
Thus, from Eq.\ (50), we have that the electromagnetic field contribution
is nearly equal to the bare mass, and about half of the measured mass,
$$
\delta m\approx 0.942 m\approx 0.485 m_m.\eqno{(55)}
$$
{}From Eq.\ (51) we obtain the critical radius $r_s$,
$$
r_s\approx 5.99\times10^{-45}\>{\rm fm},
$$
which is an ultramicro black hole indeed!  It is so small, in fact, that
we need to ask whether the black-hole radius due to the bare mass $m$
doesn't seriously compete.  The negative answer to the question is due
to the fact that spontaneous virtual pair production and annihilation
will also cause the bare mass to be slightly smeared out on the scale of
a Compton wavelength, rather than its being a point mass (this is an
aspect of the {\it zitterbewegung} effect).  For the electron at ``rest'',
Weisskopf [1] gives the following bare charge density, $\tilde\rho$,
$$
\tilde\rho (r) = -e\,d(r),\eqno{(56a)}
$$
where $d(r)$ is the spherically symmetric probability density,
$$
d(r)\equiv \int{d^3\vec k\over (2\pi)^3}\,%
{\exp\left (i\vec k\cdot\vec r\right )\over
\left (\left (|\vec k|/ m\right )^2 + 1\right )^{1\over 4}}%
\>,\eqno{(56b)}
$$
which has the small $r$ asymptotic behavior,
$$
d(r)\sim {m^{1\over 2}\over \left (32\pi^3\right )^{1\over 2}\, r^{5\over 2}}
\quad{\rm for}\; r\ll {1\over m}.\eqno{(56c)}
$$
We remark here in passing that a charge density which behaves as
$Cr^{-{5\over 2}}$ as $r\to 0$, as does that of Eqs.\ (56), produces an elect%
rostatic field whose ${\rm modulus}\sim 8\pi |C| r^{-{3\over 2}}$ as $r\to 0$,
which it turn implies an electrostatic energy
${\rm density}\sim 8\pi C^2 r^{-3}$ as $r\to 0$, rather similar to what is
found in Eq.\ (41a)---the latter, however, includes magnetic (spin) effects
as well as electrostatic ones.

In view of the bare charge density of Eqs.\ (56) for the electron at
``rest'', it is reasonable to as well model its distribution of bare
mass by the energy density,
$$
\tilde\veps (r) = m\,d(r),\eqno{(57a)}
$$
which, of course, has the small $r$ asymptotic behavior,
$$
\tilde\veps (r)\sim {m^{3\over 2}\over \left (32\pi^3\right )^{1\over 2}\,%
r^{5\over 2}}\quad{\rm for}\; r\ll {1\over m}.\eqno{(57b)}
$$
Putting Eq.\ (57b) into Eq.\ (29), we have that the bare mass distribution
critical radius $\tilde r_s$ satisfies,
$$
{m^{3\over 2}\over \left (2\pi\tilde r_s\right )^{1\over 2}} = {1\over 2G},
$$
or
$$
\tilde r_s={2G^2m^3\over\pi}={2G^2m_m\!\!\!{}^3\over\pi (1+\Delta)^3}.%
\eqno{(58)}
$$
We can thus see that $\tilde r_s\ll r_s$.  Indeed,
$$
{\tilde r_s\ovwp r_s}={8Gm_m\!\!\!{}^2\over 3\pi^2 e_m\!\!\!{}^2 (1+\Delta)^2}
\approx 1.72\times 10^{-44},\eqno{(59)}
$$
so, notwithstanding the ultrasmall radius $r_s$ of the electron's electro%
magnetic field induced black hole, the black-hole radius $\tilde r_s$ assoc%
iated with its bare mass distribution is vastly smaller, and doesn't inter%
fere with our calculation of $\delta m$.
\medskip
\centerline{{\bf IV.}\enspace{\ssbf Speculations about Future Developments}}
\smallskip
Having adduced from self-gravitational effects a finite electromagnetic
contribution, of order of the bare mass, to the measured mass of the
electron, we can ask what implications self-gravitational effects might
hold for the remaining ultraviolet divergences of quantum electrodynamics.
The key remaining one is associated with the photon's virtual pair dissoci%
ation and recombination Feynman diagram of Fig.\ (3a) (usually referred
to as ``the vacuum polarization diagram'' despite the fact that such
an appellation may be more appropriate to the diagram of Fig.\ (2a)), which
diverges logarithmically and makes an infinite contribution to charge re%
normalization [8].  We recall, of course, that Eq.\ (41b) implies that the
spontaneous vacuum virtual pair and photon production diagram of Fig.\ (2a)
makes a {\it finite\/} (but not small!) contribution to charge renormaliza%
tion.

We now proceed to speculate about the possibility of gravitationally
resolving the ultraviolet divergence in the diagram of Fig.\ (3a).
If we remove one of the external photon propagators from Fig.\ (3a), we
produce Fig.\ (3b), which lends itself to the interpretation of being the
four-dimensional Fourier transform of the stress-energy contribution to
the electromagnetic field of the photon's dissociation into and recombina%
tion from the virtual pair.  This subdiagram carries the logarithmic diver%
gence of its parent diagram, but it is also a second-rank tensor object
which is symmetric in its two indices and has vanishing four-divergence
(in the algebraic sense which applies to Fourier transforms), because of
the gauge character of electromagnetism [8].  These properties, along
with dimensional consistency and physical plausibility, support the
above-stated interpretation of Fig.\ (3b) as being the four-dimensional
Fourier transform of the stress-energy contribution to the electromagnetic
field due to the photon's dissociation into and recombination from the
virtual pair.  If we now multiply this subdiagram of Fig.\ (3b) by
$k^2(2em)^{-2}$, we sever the remaining external photon propaga%
tor and remove all charge dependence, arriving at the lesser subdiagram depic%
ted in Fig.\ (3c), which still carries the logarithmic divergence and has
all the properties required of a four-dimensional Fourier transform of a
stress-energy tensor---we can only interpret it as the four-dimensional
Fourier transform of the stress-energy of the transient virtual pair {\it it%
self}.  This Fig.\ (3c) subdiagram is written as the trace of a (divergent)
convolution of two Dirac matrix functions.  These matrix functions are not
themselves divergent (only their convolution fails to converge), and each may
be individually inverse Fourier transformed back to configuration space.
Then the convolution theorem may be applied to easily obtain the
stress-energy tensor of which the diagram of Fig.\ (3c) is the Fourier
transform.  This stress-energy can be expected to be very localized in
space-time (as befits the stress-energy of a virtual pair), and the
logarithmic divergence of its Fourier transform must be reflected as
a nonintegrable local singularity in this stress-energy itself.  We may
then expect, in analogy to what we found in Section III, that self-gravi%
tational effects will, via black-hole phenomena, cut off the local singu%
larity of this stress-energy, producing a modified one which is fully
integrable and Fourier transformable, enabling us to successively remove
the logarithmic divergence from Figs.\ (3c), (3b), (3a), and charge
renormalization.

Let us now wildly leap to the happy conclusion that every ultraviolet diver%
gence of not just quantum electrodynamics, but of any quantum field theory,
elicits a self-gravitational response which renders it finite.  How would
this affect the conventional wisdom that the only admissible quantum field
theories are the renormalizable ones?  Nonrenormalizable theories would no
longer have {\it infinities\/} ultimately contaminating every quantity of
interest, instead the {\it gravitational constant} $G$ would ultimately
enter into every quantity of interest.  For, say, every strong interaction
scattering cross section to involve $G$ hardly seems physically sensible.
So the injunction against nonrenormalizable quantum field theories would
still hold, with the justification for it shifted from sheer calculational
necessity to considerations of physical plausibility.  There is one tran%
scendent exception to the renormalizability requirement under this
scenario, however, and that is quantized gravity theory itself---there can
certainly be no objection to $G$ entering into every quantity of interest
in gravity theory.  This is quite pleasing, as quantized gravity theory is
indeed nonrenormalizable in the context of the standard perturbation expan%
sion in nonnegative powers of $G$, and it would be bewildering indeed if
this had to be regarded as an insuperable pathology, given the great suc-
cess and profound physical elegance of general relativity.  However, the
indications of this paper are that the gravitational cures for quantum
field theoretic ultraviolet divergences lie with black-hole
phenomena---ultrastrong gravitational effects which involve the likes of
$G^{-1}$ and are absolutely unamenable to sensible perturbative expansion
in nonnegative powers of $G$.  Ironically, it seems that the only acceptable
{\it nonrenormalizable\/} quantum field theory, quantized gravitational
theory, is also the only one which is, in fact, self-consistently finite
{\it in its own right}.  To demonstrate this beyond doubt would, of course,
require the development of a powerful {\it nonperturbative\/} approach to
quantized gravitation.

Meantime, a program that would progressively banish the ultraviolet
divergences from renormalizable quantum field theories by the proper
invocation of strong self-gravitational effects could well stimulate
useful calculational progress in some of these theories.  For the strong
interactions, for example, the manifestly covariant perturbation expan%
sion---the environment in which the egg-walk of the renormalization
program is carried out---may be calculationally quite inappropriate be%
cause of slow or nonconvergence.  One would like to have
{\it nonperturbative\/} approximation techniques to deal with such
strongly coupled renormalizable theories, but the development of
these has been hindered by uncertainty over whether the ultraviolet
divergences are properly dealt with.  Physical understanding that the
proper introduction of self-gravitational effects in fact eliminates
these divergences could well encourage the confident development of
more powerful approximation techniques for such important classes of
quantum field theories.
\vfill\eject
\centerline{{\ssbf Acknowledgment}}
\smallskip
The author wishes to thank Theodore Garavaglia, Samir Dutt, Barry Newberger,
James Ellison, and Narayan Mahale, all former colleagues at the Supercon%
ducting Super Collider Laboratory, for their aid, suggestions, and useful
conversations.
\bigskip
\medskip
\centerline{{\ssbf References}}
\medskip
{\parindent=2em \parskip=1ex
\item{1.} V. F. Weisskopf, {\it Phys. Rev.\/} {\bf 56}, 72 (1939), p. 77.
\item{2.} S. Weinberg, {\it Gravitation and Cosmology: Principles and
          Applications of the General Theory of Relativity\/} (John Wiley
          \& Sons, New York, 1972), pp. 299--302.
\item{3.} J. D. Bjorken and S. D. Drell, {\it Relativistic Quantum
          Mechanics\/} (McGraw-Hill, New York, 1964), p. 162.
\item{4.} Weinberg, pp. 175--178, {\it op. cit.}
\item{5.} Weinberg, pp. 77--79 and p. 152, {\it op. cit.}
\item{6.} Weinberg, pp. 165--171, {\it op. cit.}
\item{7.} Bjorken and Drell, p. 284, {\it op. cit.}
\item{8.} Bjorken and Drell, pp. 153--161, {\it op. cit.}
\par}
\vfill\eject
\null\bigskip\bigskip
\centerline{{\ssbf Figure Captions}}
\bigskip
{\parindent=8em \parskip=4ex
\item{Fig.\ 1.} Feynman diagram for the electromagnetic mass contribution to
                the electron.
\item{Fig.\ 2.} Time orderings of the Feynman diagram for the electromagnetic
                mass contribution to the electron.  (a) Spontaneous vacuum
                virtual pair and photon production.  (b) Virtual photon
                emission and reabsorption.
\item{Fig.\ 3.} Feynman diagrams for the dissociation of photons into,
                and recombination from, virtual pairs.  (a) Contribution to
                the photon propagator.  (b) Subdiagram which is the Fourier
                transform of the contribution to the electromagnetic
                field stress-energy from this pair dissociation process.
                (c) Subdiagram which is the Fourier transform of the
                stress-energy of the transient virtual pair itself.
\par}
\bye